%% file: main.tex
\documentclass[final,5p,times,twocolumn]{elsarticle}
\usepackage{amsmath,amsfonts,amsthm,amssymb}
\usepackage[usenames,dvipsnames]{color}
\definecolor{darkblue}{RGB}{0,0,196}
\definecolor{darkgreen}{RGB}{0,120,0}
\usepackage[colorlinks=true,linktocpage=true,linkcolor=darkblue,citecolor=red,urlcolor=darkblue]{hyperref}
\usepackage{cancel}
\usepackage{bbold}
\usepackage{multirow}
\usepackage{longtable}
\usepackage{color}
\usepackage[normalem]{ulem}
\usepackage{hyperref}
\usepackage{bigints}
\usepackage{xparse}
\usepackage{physics}
\usepackage{verbatim}
\usepackage{minibox}
\usepackage{comment}
\usepackage{appendix}
\usepackage{marginnote}
\usepackage{graphicx}
\usepackage[nice]{nicefrac}
\usepackage{soul}

\usepackage{stackengine}

\newcommand\oU[1]{\ensurestackMath{\stackon[1pt]{#1}{\mkern2mu\bullet}}}
\newcommand\oX[1]{\ensurestackMath{\stackon[1pt]{#1}{\mkern2mu\star}}}
\newcommand\oY[1]{\ensurestackMath{\stackon[1pt]{#1}{\mkern2mu\smwhitestar}}}
\newcommand\oZ[1]{\ensurestackMath{\stackon[1pt]{#1}{\mkern2mu\circ}}}
\def\HP{\hphantom{\alpha}}
\usepackage{amsmath}
\usepackage{hepunits}

\newcommand{\FUG}{\ensuremath{\xi}}

\def\FUGM{\a_M}
\def\GLW{{\rm GLW}}
\newcommand{\Inqr}[3]{I_{#1#2}^{(#3)}}

\journal{Nuclear Physics A}

\begin{document}
\include{commands}
\begin{frontmatter}
 
\title{Relativistic hydrodynamics with spin in the presence of electromagnetic fields}

%
\author[inst1,inst2,inst3]{Rajeev Singh \corref{cor1}}
\cortext[cor1]{Corresponding author}
\ead{rajeev.singh@stonybrook.edu}
\affiliation[inst1]{Institute  of  Nuclear  Physics  Polish  Academy  of  Sciences,  PL-31-342  Krakow,  Poland}
\affiliation[inst2]{Center for Nuclear Theory, Department of Physics and Astronomy, Stony Brook University, Stony Brook, New York, 11794-3800, USA}

\author[inst3]{Masoud Shokri}
\ead{shokri@itp.uni-frankfurt.de}
\affiliation[inst3]{Institute for Theoretical Physics, Goethe University Frankfurt,
Max-von-Laue-Strasse 1, D-60438 Frankfurt am Main, Germany}

\author[inst4]{S.~M.~A.~Tabatabaee Mehr}
\ead{tabatabaee@ipm.ir}
\affiliation[inst4]{School of Particles and Accelerators, Institute for Research in Fundamental Sciences (IPM), P.O. Box 19395-5531, Tehran, Iran}
\begin{abstract}
We extend the classical phase-space distribution function to include the spin and electromagnetic fields coupling and derive the modified constitutive relations for charge current, energy-momentum tensor, and spin tensor. Because of the coupling, the new tensors receive corrections to their perfect fluid counterparts and make the background and spin fluid equations of motion communicate with each other. We investigate special cases which are relevant in high-energy heavy-ion collisions, including baryon free matter and large mass limit. Using Bjorken symmetries, we find that spin polarization increases with increasing magnetic field for an initially positive baryon chemical potential. The corrections derived in this framework may help to explain the splitting observed in Lambda hyperons spin polarization measurements.
\end{abstract}
\end{frontmatter}
%

\section{Introduction}
\label{sec:introduction}
%
 Over the last few decades, relativistic heavy-ion collision experiments at RHIC and LHC have provided a unique opportunity for the study of the properties of hot and dense relativistic nuclear matter under extreme conditions~\cite{Florkowski:1321594}. The system produced by the colliding nuclei at high collision energies has been shown to quickly evolve from the initial nonequilibrium glasma state to an equilibrated quark-gluon plasma (QGP) phase and eventually recombine into hadrons below the freeze-out temperature~\cite{Kharzeev:2000ph,Heinz:2001xi,Gyulassy:2004zy,Shuryak:2004cy}. Using the theory of relativistic hydrodynamics it was shown that QGP matter produced this way forms the smallest fluid droplets ever observed which exhibit a nearly perfect fluid behavior~\cite{Romatschke:2007mq, Heinz:2013th,Son:2006em,Schafer:2009dj,Schenke:2021mxx}. The success of hydrodynamic description of relativistic heavy-ion collisions has opened new avenues in theoretical studies of relativistic matter, including the development of the theory of hydrodynamics and its off equilibrium applications~\cite{Florkowski:2017olj}.
 Particularly interesting directions of study which are recently being followed include investigations of spin polarization phenomena in relativistic nuclear matter~\cite{Becattini:2020ngo,lisa2021} and its dynamics under the influence of electromagnetic fields (EM)~\cite{Kharzeev:2013jha,Hattori:2016emy,Hattori:2022hyo}. Recent spin polarization measurements of emitted particles provided a new probe for studying QGP~\cite{STAR:2017ckg,Adam:2018ivw,STAR:2019erd,ALICE:2019aid,Acharya:2019ryw,Kornas:2019,STAR:2021beb,ALICE:2021pzu} and triggered many theoretical developments~\cite{Florkowski:2017ruc,Florkowski:2017dyn,Florkowski:2018ahw,Florkowski:2018fap,Florkowski:2019qdp,Singh:2020rht,Singh:2021man,Florkowski:2021wvk,Bhadury:2020puc,Bhadury:2020cop}, see also Refs.~\cite{Hattori:2019lfp,Fukushima:2020ucl,Li:2020eon,Montenegro:2020paq,Weickgenannt:2020aaf,Garbiso:2020puw,Gallegos:2021bzp,Sheng:2021kfc,Speranza:2020ilk,Bhadury:2021oat}. Being oriented along the direction of the global angular momentum of the colliding system, microscopically, the spin polarization is believed to arise because of the spin-orbit coupling~\cite{Liang:2004ph}. On the other hand, since at the macroscopic level, the QGP admits close-to-equilibrium dynamics, it is believed that the spin degrees of freedom undergo the thermalization as well. This may allow generation of spin polarization through the coupling between the fluid vorticity and spin of its constituents~\cite{Becattini:2013fla}.
While the agreement between the global polarization measurements and ``spin-thermal'' models supports the hypothesis of polarization-vorticity coupling~\cite{Becattini:2016gvu,Karpenko:2016jyx,Pang:2016igs,Xie:2017upb,Becattini:2017gcx,Fu:2020oxj}, the same theories do not quite explain the measurements related to differential observables~\cite{Karpenko:2016jyx,Becattini:2017gcx,Florkowski:2019voj}; though there are some recent advances in this direction~\cite{Becattini:2021suc,Becattini:2021iol,Fu:2021pok,Florkowski:2021xvy}. 
 Discrepancies between the theoretical predictions and the experimental data indicate that the current theoretical understanding of spin polarization dynamics in heavy-ion collisions is incomplete. If the spin degrees of freedom are thermalized, they should be incorporated into the hydrodynamic formalism on the same footing as the other macroscopic quantities, allowing for their non-trivial dynamics.
To this end, several frameworks considering spin as a dynamical quantity were developed that used thermodynamic equilibrium~\cite{Becattini:2009wh},
perfect fluid spin hydrodynamics~\cite{Florkowski:2017ruc,Florkowski:2017dyn},
effective action approach~\cite{Montenegro:2018bcf,Montenegro:2020paq,Gallegos:2021bzp,Serenone:2021zef,Torrieri:2022xil}, method of entropy current analysis~\cite{Hattori:2019lfp,Fukushima:2020ucl,Li:2020eon,She:2021lhe,Daher:2022xon,Cao:2022aku}, statistical operator formalism~\cite{Hu:2021lnx}, nonlocal collisions~\cite{Hidaka:2018ekt,Yang:2020hri,Wang:2020pej,Weickgenannt:2020aaf,Weickgenannt:2021cuo,Sheng:2021kfc,Hu:2021pwh,Hu:2022lpi,Das:2022azr,Weickgenannt:2022zxs}, kinetic theory for massless fermions~\cite{Stephanov:2012ki,Chen:2014cla,Gorbar:2017toh,Hidaka:2018ekt,Shi:2020htn}, holographic method~\cite{Heller:2020hnq,Gallegos:2020otk,Garbiso:2020puw,Gallegos:2021bzp,Hongo:2021ona,Gallegos:2022jow}, anomalous hydrodynamics~\cite{Son:2009tf,Kharzeev:2010gr}, and Lagrangian method~\cite{Montenegro:2017rbu,Montenegro:2017lvf}.
Also see Refs.~\cite{Ambrus:2019khr,Ambrus:2019ayb,Ambrus:2020oiw,Becattini:2020xbh,Gao:2021rom,Yi:2022nnc} for the studies related to helicity polarization.
Relativistic hydrodynamics with spin proposed in Refs.~\cite{Florkowski:2017ruc,Florkowski:2017dyn} was further developed~\cite{Florkowski:2018ahw,Florkowski:2018fap,Florkowski:2019qdp,Singh:2020rht,Singh:2021man,Florkowski:2021wvk} and also extended to dissipative systems in Refs.~\cite{Bhadury:2020puc,Bhadury:2020cop}.
In its current form, the spin hydrodynamics framework does not include interactions with the electromagnetic field which may possibly be present in the early-time evolution of QGP. Such a coupling may be crucial for the explanation of the splitting in $\Lambda$ and $\Bar{\Lambda}$ spin polarization signal
observed in the experiments~\cite{STAR:2017ckg,Li:2021zwq}, which may arise because of opposite magnetic moments of these particles \cite{Becattini:2016gvu}, for other approaches, see Refs.~\cite{Vitiuk:2019rfv,Ambrus:2020oiw}.
 In this work, we extend the spin hydrodynamics framework developed in Refs.~\cite{Florkowski:2018ahw,Florkowski:2018fap,Florkowski:2019qdp} and include the coupling between spin and electromagnetic fields in the phase-space distribution function of the constituent particles. We obtain modified constitutive relations for the baryon charge current, energy-momentum tensor, and spin tensor arising because of the coupling. Then, we investigate special cases of these quantities for the baryon-free system and the large mass limit, which are relevant to the physics of ultra-relativistic heavy-ion collisions. Finally, we study equations of motion in the case of Bjorken symmetry in the ideal MHD limit and obtain the evolution of the spin component.
Although the qualitative features of spin polarization remains same as in Ref.~\cite{Florkowski:2019qdp} due to strong symmetries assumed herein, however,
we find that magnetic field positively enhances the spin polarization for an initial positive baryon chemical potential.
We think that in more realistic setup our framework will prove to be crucial in understanding the splitting between $\Lambda$ and $\Bar{\Lambda}$ spin polarization.
%
\section{Conventions}
\label{sec:convention}
%
We use the mostly-minus Minkowski metric signature which reads $g_{\mu\nu}={\rm diag}\left(+1,-1,-1,-1\right)$. As a result, the fluid four-velocity is normalized as $U^\mu U_\mu = 1$. The operator $\Delta^{\mu\nu}=g^{\mu\nu}-U^\mu U^\nu$ projects tensors on the space transverse to $U^\mu$. A tensor $M_{\mu\nu}$ can be decomposed into a symmetric $M_{(\mu\nu)}\equiv\half\left(M_{\mu\nu}+M_{\nu\mu}\right)$ and an asymmetric $M_{[\mu\nu]}\equiv\half\left(M_{\mu\nu}-M_{\nu\mu}\right)$ part. We denote Levi-Civita symbol as $\epsilon^{\a\b\g\d}$ which is totally antisymmetric and use the convention of $\epsilon^{0123}=-\epsilon_{0123}=1$. Euclidean three-vectors are denoted with boldface, like $\vb{B}$, as opposed to four-vectors. For the scalar and Frobenius product we use the notation $a \cdot b \equiv a^\mu b_\nu$ and $A : B \equiv A^{\mu\nu}B_{\mu\nu}$, respectively. Throughout the paper we assume natural units {\it i.e.} $c = \hbar = k_B~=1$, unless stated otherwise.
%
\section{One-particle distribution function: coupling spin to EM fields}
\label{sec:HydroEM}
%
In this section we extend the classical phase-space spin distribution function~\cite{Florkowski:2018fap,Bhadury:2020puc} by introducing a term coupling the spin to the external EM fields, in a way suggested in Ref.~\cite{Becattini:2020ngo}.

For the classical treatment of particles having spin one half and mass $m$, phase-space distribution function with spin reads~\cite{Florkowski:2018fap,Bhadury:2020puc}
 \begin{equation}
f^\pm_{0}(x,p,s) = f^\pm_{0}(x,p)\exp[\half\,\omega(x) :  s(p)]\,,
\label{eq:fpm-spin}
 \end{equation}
where $\omega_{\alpha \beta}(x)$ is the spin polarization tensor and $s^{\alpha\beta}(p)$ represents the particle internal angular momentum expressed with spin four-vector $s^{\alpha}$ and four-momentum $p^{\alpha}$ as
\bea
s^{\alpha\beta} = \f{1}{m} \epsUabgd p_\gamma s_\delta\,.
\label{eq:salbe}
\eea
In \rf{eq:fpm-spin}, $f_{0}^{\pm}(x,p)=\exp[-p \cdot \beta(x)\pm\xi(x)]$ is the J\"uttner distribution, with $\xi(x)$ being the ratio of the baryon chemical potential $\mu(x)$ over temperature $T(x)$, $\xi=\mu/T$, and $\beta_{\mu}(x)$ is the ratio of the fluid four-velocity $U_\mu(x)$ to temperature, $\beta_\mu=U_\mu/T$.
Note that, the classical distribution function~\eqref{eq:fpm-spin} is valid only for the case of local collisions between particles, however, it can be extended to include nonlocal effects through the gradients of $f^\pm_{0}(x,p,s)$~\cite{Florkowski:2018fap}.

We generalize the phase-space distribution function \rfn{eq:fpm-spin} to a case of interaction between the particle magnetic moment and the external EM field by introducing the modified distribution function in the form
\begin{equation}
	f^\pm_{\rm s} (x,p,s) = f^\pm_{\rm 0}(x,p,s)\exp\left[\mp\alpha_M(x) F(x) : s\right],
	\label{eq:fpm-em-spin}
\end{equation}
where $F_{\mu\nu}$ is the Faraday tensor expressed in terms of electric $E_\mu$ and magnetic $B_\mu$ four-vectors as
\begin{equation}
	F_{\mu\nu} = E_\mu U_\nu -E_\nu U_\mu + \epsLmnab U^\a B^\b,
\end{equation}
with
\begin{equation}
    E^\mu \equiv F^{\mu\nu}U_\nu\,,\qquad B^\mu \equiv \half \epsUmnab F_{\nu\a} U_\b\,.
\end{equation}
In \rf{eq:fpm-em-spin}, $\alpha_M = \mu_{M}/T$ where $\mu_{M} = g_{Q} \mu_{N}$ is the magnetic moment of the quasiparticles and $\mu_{N}$ being the nuclear magneton. In this work, for simplicity, we assume that the quasiparticles are  $\Lambda$ hyperons, with $g_\Lambda = -0.6138 \pm 0.0047$~\cite{ParticleDataGroup:2020ssz}. However, it should be mentioned that, a more realistic setup needs multiple quark-like quasiparticles, with constitutive masses, see for example~\cite{Singh:2021man}. 
Keeping in mind the smallness of the amplitude of spin polarization in measurements~\cite{STAR:2017ckg},
we assume the small polarization limit~\cite{Florkowski:2019qdp}, $\omega_{\alpha\beta} \ll 1$.
We also assume weak EM fields $eB \ll M^2$, with $M$ being a typical energy scale relevant to the system in consideration. If the quasiparticles are assumed to be electrically neutral Lambda hyperons, then $M \sim m_\Lambda$. The aforementioned assumptions allow \rf{eq:fpm-em-spin} to be approximated as
\begin{eqnarray}
		f^\pm_{\rm s}(x,p,s) &=& f_{0}^{\pm}(x,p)\left(1 +\f{1}{2}  \omega : s\right)
	\left(1 \mp \FUGM F : s\right),
		~~~~\label{feq}
	\end{eqnarray}
where $f_{\rm s}^+(f_{\rm s}^-)$ represents the distribution function for particles (antiparticles). 
Similarly to the Faraday tensor, the spin polarization tensor $\omega_{\alpha \beta}$ in \rf{feq} can be decomposed with respect to the fluid four-velocity as~\cite{Florkowski:2021wvk}
 \begin{equation}
\omega_{\mu\nu} = \kappa_\mu U_\nu - \kappa_\nu U_\mu + \epsilon_{\mu\nu\a\b} U^\a \omega^{\b}, \lab{spinpol1}
 \end{equation}
where the four-vectors $\kappa_\mu$ and $\omega_\mu$,
 \begin{equation}
\kappa_\mu= \omega_{\mu\a} U^\a, \quad \omega_\mu = \half \epsilon_{\mu\a\b\gamma} \omega^{\a\b} U^\gamma, \label{eq:kappaomega}
 \end{equation}
are orthogonal to $U^\mu$~\cite{Florkowski:2021wvk}.
The above conditions leave $\kappa_\mu$ and $\omega_\mu$ with three degrees of freedom each, constituting together the same number of degrees of freedom as $\omega_{\mu\nu}$.
These four-vectors can be written in terms of three space-like orthonormal vectors $X^\mu$, $Y^\mu$, and $Z^\mu$ which span the plane transverse to $U^\mu$. These vectors, together with $U^\mu$, form a basis satisfying~\cite{Florkowski:2019qdp,Florkowski:2021wvk}
 \begin{equation}
U \cdot U = 1\,, \quad X \cdot X = Y \cdot Y = Z \cdot Z = -1\,.
\label{eq:four-vectors}
 \end{equation}
Consequently,
 \begin{align}
\kappa^\a &=  C_{\kappa X} X^\a + C_{\kappa Y} Y^\a + C_{\kappa Z} Z^\a\,, \lab{eq:k_decom}\\
\omega^\a &=  C_{\omega X} X^\a + C_{\omega Y} Y^\a + C_{\omega Z} Z^\a\,. \lab{eq:o_decom}
 \end{align}
Here, $C_{\boldsymbol{\kappa}} = (C_{\kappa X}, C_{\kappa Y}, C_{\kappa Z})$, and
$C_{\boldsymbol{\omega}} = (C_{\omega X}, C_{\omega Y}, C_{\omega Z})$ will be referred as the spin polarization components.
\section{Constitutive relations in the presence of EM fields}
\label{sec:cons_relations}
In the following we derive the constitutive relations for the baryon charge current, energy-momentum tensor, and spin tensor from the distribution function~\eqref{feq} introduced in the previous section.
\subsection{Charge current}
\label{subsubsec:1}
The baryon charge current is the first moment of the modified distribution function (\ref{feq})~\mbox{\cite{Florkowski:2018fap,Florkowski:2018ahw}}
 \begin{equation}
N^\lambda 
= \!\int \! \mathrm{dP}~\mathrm{dS} \, \, p^\lambda \, \left[f^+_{\rm s}(x,p,s)\!-\!f^-_{\rm s}(x,p,s) \right],
\label{eq:Neq-sp0}
 \end{equation}
where the invariant integration measures for momentum $\mathrm{dP}$ and spin $\mathrm{dS}$ are defined, respectively, as~\cite{Florkowski:2018fap}
 \begin{align}
\mathrm{dP} &= \frac{\mathrm{d}^3p}{(2 \pi )^3 E_p}\,,\nn\\ \mathrm{dS} &= \frac{m}{\pi {\mathfrak{s}}} \, \mathrm{d}^4s~\delta(s\cdot s + {{\mathfrak{s}}}^2)~\delta(p\cdot s)\,.
\label{eq:dS}
 \end{align}
Here $E_p$ is the particle energy and
$\mathfrak{s}^2$ is the length of the spin vector, which for spin half particles equals 3/4~\cite{Florkowski:2018fap}.

Similarly to the current of a dissipative charged fluid~\cite{Kovtun:2012rj}, $N^\lambda$ can be decomposed as 
 \begin{equation}
N^{\lambda} =  {\cal N} U^{\lam} + N_\perp^\lam = \left({\cal N}_{\rm PF}+{\cal N}_{\rm EM}\right) U^{\lam} + N_\perp^\lam \,,
\label{eq:chargecurrent}
 \end{equation}
where $\cN_{\rm PF}$ is the perfect fluid baryon charge density~\cite{Florkowski:2021wvk}
\ba
{\cal N}_{\rm PF} = 4 \, \sinh(\xi)\, {\cal N}_{(0)}\,.
\lab{nden}
\ea
The factor $4 \, \sinh(\xi)$ accounts for spin and particle-antiparticle degeneracies, and ${\cal N}_{(0)}$
is number density of the spinless neutral classical massive particles~\cite{Florkowski:1321594}
 \begin{equation} 
{\cal N}_{(0)} = gT^3\,z^2   K_{2}(z)\,, \label{N0}
 \end{equation}
with $g = 1/(2\pi^2)$,
$z=m/T$ and $K_n$ being $n^{th}$
modified Bessel function of second kind.
In \rf{eq:chargecurrent}, ${\cal N}_{\rm EM}$ is the charge density modification and $N_\perp^\lam$ is the transverse current, both due to spin-EM coupling. They read
\ba
{\cal N}_{\rm EM} &=& \alpha_M  \cosh(\xi){\cal N}_{(0)} \epsilon^{\beta\gamma\nu\mu} \omega_{\beta\gamma} U_\mu B_\nu \,,\nn\\
N_\perp^\lam &=& \alpha_M  \cosh(\xi) \mathcal{A}_3 \big(U^\lam F^{\beta\gamma} + 6 U^\lam U^{[\beta} E^{\gamma]} \nn\\
&&- U^{[\beta} F^{\gamma]\lam} - g^{\lambda
[\beta} E^{\gamma]}\big)\omega_{\beta\gamma}\,,
\label{eq:couplingNden}
\ea
where
$
{\cal A}_3 = -\left(2\left({\cal E}_{(0)} + {\cal P}_{(0)}\right)\right)/(T\, z^2)\, 
$~\cite{Florkowski:2017ruc,Florkowski:2018fap,Florkowski:2018ahw,Florkowski:2021wvk} with ${\cal P}_{(0)}$, and ${\cal E}_{(0)}$ denoting the pressure, and energy density of the spinless and neutral classical massive particles, defined as~\cite{Florkowski:1321594}
 \begin{equation}
{\cal P}_{(0)} = {\cal N}_{(0)} T\,, \qquad
{\cal E}_{(0)} =  g\, z^3 T^4 K_{1}(z) + 3 {\cal P}_{(0)}\,, \label{P0}
 \end{equation}
respectively. 

The conservation of charge current, $\partial_\mu N^\mu = 0$, leads to the first equation of motion
 \begin{equation}
\UD {\mathcal{N}}_{\rm PF} + \UD{\mathcal{N}}_{\rm EM} + \left(\mathcal{N}_{\rm PF} + \mathcal{N}_{\rm EM}\right) \theta_U  = - \partial \cdot N_\perp\,,
\label{eq:NEoMgeneralMS}
 \end{equation}
wherein $\UD {\cdots}\equiv U \cdot\p\,{\cdots}$ denotes the co-moving temporal derivative~\cite{Romatschke:2017ejr} and $\theta_U\equiv\p \cdot U$ is the expansion scalar. If the quasi-particles carry a electric charge $q$, the electric current is 
\begin{equation}
J^\mu=q \, N^\mu\,,
\label{eq:J}
\end{equation}
resulting in a ``back-reaction'' of spin-EM coupling with EM fields via Maxwell equations. Since we are taking Lambda hyperons, which are electrically neutral, as the quasiparticles of the fluid the electric current vanishes~\cite{Denicol:2019iyh}.
%
\subsection{Energy-momentum tensor}
\label{sec:emtensor}
%
The energy-momentum tensor is the second moment of the distribution function (\ref{feq}), namely~\cite{Florkowski:2018fap,Florkowski:2018ahw}
\bea
T^{\mu \nu}
&=& \int \mathrm{dP}~\mathrm{dS} \, \, p^\mu p^\nu \, \left[f^+_{\rm s}(x,p,s) + f^-_{\rm s}(x,p,s) \right].~~
\label{eq:Teq-sp02}
\eea
Performing the spin integration followed by the momentum integration one obtains
 \begin{equation}
T^{\mu \nu} = \cE U^\mu U^\nu - \cP \Delta^{\mu\nu} + \cQ^\mu U^\nu+ \cQ^\nu U^\mu +\cT^{\mu\nu}\,.
\label{eq:TMU1}
 \end{equation}
Here $\cE$ is the modified energy density
\begin{eqnarray}
\cE &\equiv& U_\mu U_\nu T^{\mu \nu} =
\mathcal{E}_{\rm PF} + \mathcal{E}_{\rm EM}\,,
\label{eq:cE}
\end{eqnarray}
with
 \begin{align}
{\cal E}_{\rm PF} &= 4 \cosh(\xi) {\cal E}_{(0)}\,,\label{eq:edensity}\\
{\cal E}_{\rm EM} &= \alpha_M
\sinh(\xi)\bigg\{ {\cal E}_{(0)} \,\omega : F\,   \nn\\
&+ 2 \Big[\left(\Inqr{4}{0}{0}+\Inqr{4}{1}{0}\right)\kappa\cdot E -2\Inqr{4}{1}{0}\omega\cdot B\Big]\bigg\}\,,\nn
 \end{align}
and $\cP$ is the modified pressure
\begin{eqnarray}
\cP &\equiv& -\frac{1}{3}\Delta : T =  \mathcal{P}_{\rm PF} + \mathcal{P}_{\rm EM}\,,
			\label{eq:cP}
\end{eqnarray}
with
 \begin{align}
{\cal P}_{\rm PF} &= 4 \cosh(\xi) \, {\cal P}_{(0)}\,, \label{eq:pressure}\\
{\cal P}_{\rm EM} &= \alpha_M
\sinh(\xi) \bigg\{{\cal P}_{(0)} \,\omega : F \nn\\
&- 2 \Big[\left(\Inqr{4}{1}{0}+\frac{5}{3}\Inqr{4}{2}{0}\right)\kappa\cdot E
-\frac{10}{3}\Inqr{4}{2}{0}\omega\cdot B\Big]\bigg\}.\nn
 \end{align}
In \rf{eq:TMU1}, $\cQ^\mu$ is a transverse vector current that resembles the heat current, and $\cT^{\mu\nu}$ is a transverse traceless tensor similar to the stress tensor of a dissipative fluid~\cite{Kovtun:2012rj} given by
\begin{eqnarray}
			\cQ^\mu &\equiv& \Delta^{\mu}_{\HP\a}U_\b T^{\a\b}\nn\\
			&=& 2\,\FUGM \sinh(\FUG) \Inqr{4}{1}{0}\epsUmnab\,U_\nu\left(E_\alpha \omega_\beta-B_\alpha \kappa_\beta \right),
			 \label{eq:cQ}
			 \\
			 \cT^{\mu\nu} &\equiv& \Delta^{\mu\nu}_{\alpha\beta} \, T^{\a\b}
			 \nn\\
			 &=& 
			 4\,\FUGM \sinh(\FUG)\Inqr{4}{2}{0}\Big(
			 E^{(\mu}\kappa^{\nu)}+ B^{(\mu}\omega^{\nu)}\nn\\
			 & & -
		 \inv{3}\Delta^{\mu\nu}\left(\kappa\cdot E+\omega \cdot B\right)
			 \Big)\,.
			 \label{eq:cT}
\end{eqnarray}
where $\Delta^{\mu\nu}_{\alpha\beta} \equiv (1/2)\left[\Delta^\mu_{\HP\a}\Delta^\nu_{\HP\b}+\Delta^\nu_{\HP\a}\Delta^\mu_{\HP\b}-(2/3)\Delta^{\mu\nu}\Delta_{\alpha\beta}\right]$.
The thermodynamic integrals $\Inqr{n}{q}{r}$ appearing in the equations above are defined in~\ref{app:identities}.

As usual, the conservation of energy and momentum, $\partial_\nu T^{\mu\nu} = F^{\mu\sigma} J_\sigma$, can be decomposed into longitudinal and transverse parts with respect to $U$. The longitudinal one, the so-called energy equation, reads 
\begin{eqnarray}
\UD{\cE} + (\cE + \cP) \, \theta_U = - q E \cdot {N_\bot } + \UD{U} \cdot \mathcal{Q}  -  \nabla \cdot\mathcal{Q}  
		+ \half\,\mathcal{T} : \sigma  \,,
		\label{eq:energyEq}
\end{eqnarray}
where $\nabla_\mu \equiv \partial_\mu - U_\mu U^\nu \partial_\nu$ and
$\s^{\mu \nu }\equiv \Delta^{\mu\nu}_{\alpha\beta} \nabla^\alpha U^\beta$.
In the direction transverse to $U$ we have
\begin{eqnarray}
	&(\mathcal{E} + \mathcal{P}) \, \UD{U}^{\mu }  = \nabla^{\mu } \mathcal{P}+q\left(\cN E^{\mu } + \epsilon^{\mu\alpha \beta \sigma } B_{\alpha }{N_{\bot}}_\beta \, U_{\sigma } \right)
	\label{eq:eulerEqn}\\
	&\hspace{.5cm}
	- \Big( \UD{U} \cdot \mathcal{Q}  +  \frac{1}{2}\,
	\mathcal{T} : \sigma \Big) U^{\mu } 
	- 
	\p_{\alpha }\mathcal{T}^{\mu \alpha } -  \UD{\cQ}^{\mu } - 
	\p_\a\left(U^\a\cQ^\mu\right).
	\nn
\end{eqnarray}
%
\subsection{Spin tensor}
\label{subsubsec:3}
%
The spin tensor is defined as~\cite{Florkowski:2018fap}
 \begin{equation}
S^{\lambda, \mu\nu} = \int  \mathrm{dP}~\mathrm{dS} \, \, p^\lambda \, s^{\mu \nu} 
\left[f^+_{\rm s}(x,p,s) + f^-_{\rm s}(x,p,s) \right].
\label{SGLW}
 \end{equation}
Plugging the distribution function \rfn{feq} into the above equation and integrating over the spin and momenta gives
 \begin{equation}
S^{\lambda, \mu\nu} = S^{\lambda, \mu\nu}_{\rm PF} - 2 \alpha_M \tanh(\xi) S^{\lambda, \mu\nu}_{\rm EM}\,,
\label{SGLW1}
 \end{equation}
where~\cite{Florkowski:2018fap,Florkowski:2018ahw,Florkowski:2019qdp,Florkowski:2021wvk}
 \begin{align}
S^{\a,\b\g}_{\rm PF}
&= \cosh(\xi) \Big[{\cal A}_1 \, U^\a \,  \omega^{\b\g} 
+ {\cal A}_2 \, U^\a \,  U^{[\b} \, \omega^{\g]}_{\HP\d} \, U^\d \nn\\
&+  {\cal A}_3 \, \left(U^{[\b} \, \omega^{\g]\a}  + g^{\a[\b}\, \omega^{\g]}_{\HP\d} \, U^\d \right)\Big]\,,
\nn\\
S^{\a,\b\g}_{\rm EM}
&= \cosh(\xi)\Big[ U^\a  {\cal A}_1\,  F^{\b\g} 
+ {\cal A}_2 \,U^\a U^{[\b} \, F^{\g]}_{\HP\HP\d} \, U^\d  \nn\\
&+  {\cal A}_3 \, \left(U^{[\b} \, F^{\g]\a}  + g^{\a[\b}\, F^{\g]}_{\HP\HP\d} \, U^\d \right)\Big],
\label{SEM2}
 \end{align}
with the thermodynamic coefficients defined as
 \begin{equation}
{\cal A}_1 =  {\cal N}_{(0)}-{\cal A}_3\,, \quad
{\cal A}_2 = 2 \left[{\cal A}_1 -2 {\cal A}_3\right]\, .
\label{A1A2}
 \end{equation}
As the phase-space distribution function~\eqref{feq} is valid only for the local collisions of the particles, it suggests that the orbital contribution $(L^{\alpha,\beta\gamma})$, in the total angular momentum $(J^{\alpha,\beta\gamma})$
 \begin{equation}
J^{\alpha,\beta\gamma} = L^{\alpha,\beta\gamma} + S^{\alpha,\beta\gamma} = x^\beta\,T^{\alpha\gamma} - x^\gamma \, T^{\alpha\beta} + S^{\alpha,\beta\gamma} \,,
 \end{equation}
can be eliminated and spin $(S^{\alpha,\beta\gamma})$ can be conserved independently~\cite{Florkowski:2018fap}.
Here, the energy-momentum tensor~\eqref{eq:Teq-sp02} is, by definition, symmetric. However, it may have antisymmetric contribution if we include nonlocal collisional effects~\cite{Weickgenannt:2022zxs,Das:2022azr}, which are neglected in the current study for simplicity.

Thus, neglecting nonlocal collisions we observe that the spin tensor is a conserved current
\begin{equation}
    \p_\alpha\, S^{\alpha,\beta\gamma} = 0\,.
    \label{eq:SPIN}
\end{equation}
From above we get six equations of motion for the six spin polarization components~($C_{\boldsymbol{\kappa}}$ and $C_{\boldsymbol{\omega}}$) which, in general, are coupled to each other.
\section{Special cases}
\label{sec:scases}
In this section we investigate our formalism in some special situations.
\subsection{Baryon-free system}
\label{subsec:baryon_free}
In ultra-relativistic heavy-ion collisions, the QGP can be considered approximately as baryon-free matter~\cite{Heinz:2001xi}. In the absence of the baryon chemical potential, i.e., $\mu=0$, the spin-EM coupling can only affect the charge current sector.
The spin-EM coupling results in a non-vanishing EM component of the number density $\cN_{\rm EM}$ and transverse current $N_{\bot}^\mu$ in Eq.~(\ref{eq:chargecurrent}). Furthermore, the heat current $\mathcal{Q}^\mu$ (\ref{eq:cQ}) and stress tensor $\mathcal{T}^{\mu\nu}$ (\ref{eq:cT}) vanish, and the energy-momentum tensor is reduced to its perfect fluid form
 \begin{equation}
T^{\mu \nu} = \cE_{\rm PF} U^\mu U^\nu - \cP_{\rm PF} \Delta^{\mu\nu}\,.
 \end{equation}
Similarly, the spin tensor (\ref{SGLW1}) reduces to
 \begin{equation}
S^{\alpha, \beta\gamma}_{(\mu=0)} = U^\a \left[{\cal A}_1 \omega^{\b\g} + {\cal A}_2  U^{[\b}  \kappa^{\g]}\right]
+ {\cal A}_3 \left[U^{[\b}  \omega^{\g]\a}  + g^{\a[\b}\kappa^{\g]} \right]
\label{Sbaryon}
 \end{equation}
with no spin-EM coupling terms.
\subsection{Large mass limit}
\label{subsec:massivelimit}
Another special case is when the constituents of the fluid, say hyperons, have masses much larger than the temperature, i.e. $z = m/T \gg 1$. Then, we can neglect $\cA_3$ in \rf{SEM2}, and the electric field no longer appears in the spin tensor (\ref{SGLW1}), giving
 \begin{equation}
S^{\alpha , \beta \gamma }_{z \gg 1}
= \ch(\xi){\cal N}_{(0)}U^\alpha \epsilon^{\beta\gamma\mu\nu}  U_\mu \left(\omega_\nu - 2 \alpha_M \tanh(\xi) B_\nu \right).
\label{Smassive}
 \end{equation}
\section{0+1D: Bjorken flow}
\label{sec:Bjorken_Flow}
In this section, we examine our formulation with the Bjorken setup\footnote{In the case of Bjorken expansion the basis vectors are
$U^\a = \LR \ch(\eta), 0,0, \sh(\eta) \RR$,
$X^\a = \LR 0, 1,0, 0 \RR$,
$Y^\a = \LR 0, 0,1, 0 \RR$, and
$Z^\a = \LR \sh(\eta), 0,0, \ch(\eta) \RR$,
where $\eta$ is the space-time rapidity~\cite{Florkowski:2019qdp}.} as a special situation relevant to the ultra-relativistic heavy-ion collisions.

Bjorken symmetries simplify equations of motion significantly;
the hydrodynamic variables become functions of proper time $\tau=\sqrt{t^2-z^2}$ alone, and energy-momentum conservation in the transverse direction
\rfn{eq:eulerEqn}
is trivially satisfied.
Furthermore, the heat current $\cQ^\mu$ (\ref{eq:cQ}) and the stress tensor $\cT^{\mu\nu}$ (\ref{eq:cT}) disappear.
Yet, even in this situation, the equations of motion of background and spin are highly coupled due to spin-EM contribution.
The fluid flow is affected by the spin dynamics, which is different from the previous studies~\cite{Florkowski:2019qdp,Singh:2020rht,Singh:2021man}. 

We start with the baryon charge conservation \rfn{eq:NEoMgeneralMS}. The transverse current trivially satisfies $\p_\mu N_\bot^\mu=0$ due to Bjorken symmetries. Therefore the charge conservation reduces to an equation formally similar to that of Bjorken perfect fluid~\cite{Florkowski:2019qdp},
 \begin{equation}
\dv{\cal N}{\tau} + \frac{\cal N}{\tau} =0\,.
\label{eq:BjorkenCharge}
 \end{equation}
For simplicity, we assume uncharged particles and the ideal MHD limit. Therefore, the profile of the magnetic field is~\cite{Tabatabaee:2020efb} 
 \begin{equation}
B^\mu = \alpha \cB_0 \frac{\tau_0}{\tau} Y^\mu\,,
\label{eq:magnetic_field}
 \end{equation}
with $\cB_0 = (1/e)\, m_{\pi}^2$, with $m_\pi$ being the Pion's mass, and $\alpha$ being a positive dimensionless number. At the initial proper time $\tau_0$ the initial magnetic field is $B_0 = \alpha\cB_0$. Setting $\alpha = 0$ switches off the magnetic field. Here, $Y^\mu = (0,0,1,0)$.
Due to the magnetic field profile adopted herein, Eq.~\eqref{eq:magnetic_field}, we suspect that the only spin component $C_{\omega Y}$ will be coupled to the EM fields, which is indeed the case shown below.

The spin-EM coupling occurs through the term $F : \omega = -2B_0C_{\omega Y}\tau_0/\tau$, which gives rise to
 \begin{align}
{\cal N} &= {\cal{N}}_{\rm PF}
+ 2 \, g  T^3 \, \alpha_M \, B_0  \, C_{\omega Y} \, \frac{\tau_0}{\tau} \cosh(\xi) \Big[z^2 \, K_2(z)\nn\\
&+ 2 \, z \, K_1(z) + 2 \, \left(K_2(z) + 3 K_3(z)\right) \Big]\,,
 \end{align}
where ${\cal{N}}_{\rm PF}$ is given by Eq.~\eqref{nden}.
Here $T$, $\xi$, and $C_{\omega Y}$ are unknown functions of $\tau$.

The next equation is the energy equation which reads 
 \begin{equation}
\dv{\cE}{\tau} + \frac{\cE+\cP}{\tau} = 0\,,
\label{eq:BjorkenEnergy}
 \end{equation}
where
 \begin{align}
\mathcal{E} &= g T^4 \, z^2 \left(z \, K_1(z) + 3 \,  \, K_3 (z) \right)\nn\\
&\times 2\left( 2 \cosh(\xi) +  \alpha_M \, B_0 \, C_{\omega Y} \, \frac{\tau_0}{\tau} \sinh(\xi) \right),
\label{eq:edenSolnMS}\\
\mathcal{P} &= {\cal{N}}_{\rm PF} T \left( \coth{(\xi)} + \alpha_M \,
B_0 \, C_{\omega Y} \, \frac{\tau_0}{2\tau}\right).
\label{eq:pdenSolnMS}
 \end{align}
\begin{figure}[!ht]
\centering
\includegraphics[angle=0,width=\columnwidth]{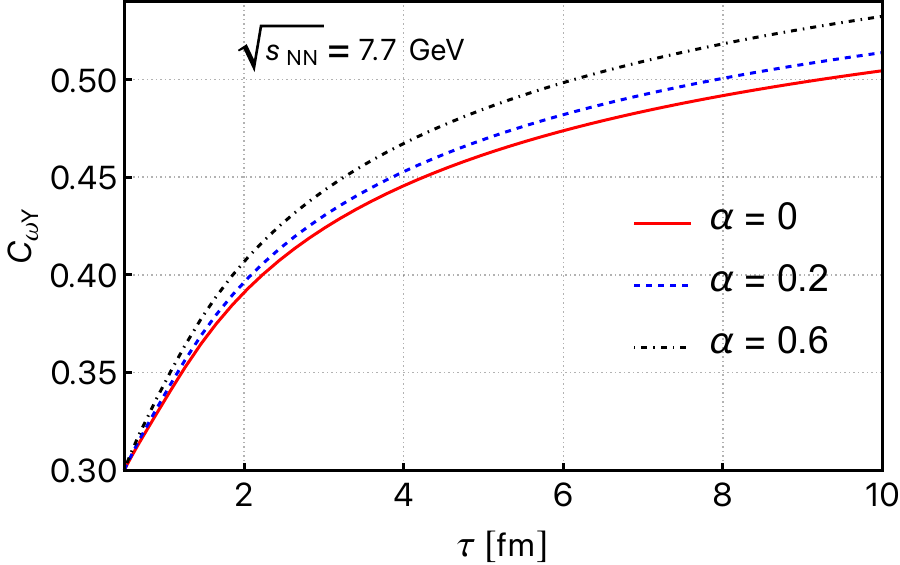}
\includegraphics[angle=0,width=\columnwidth]{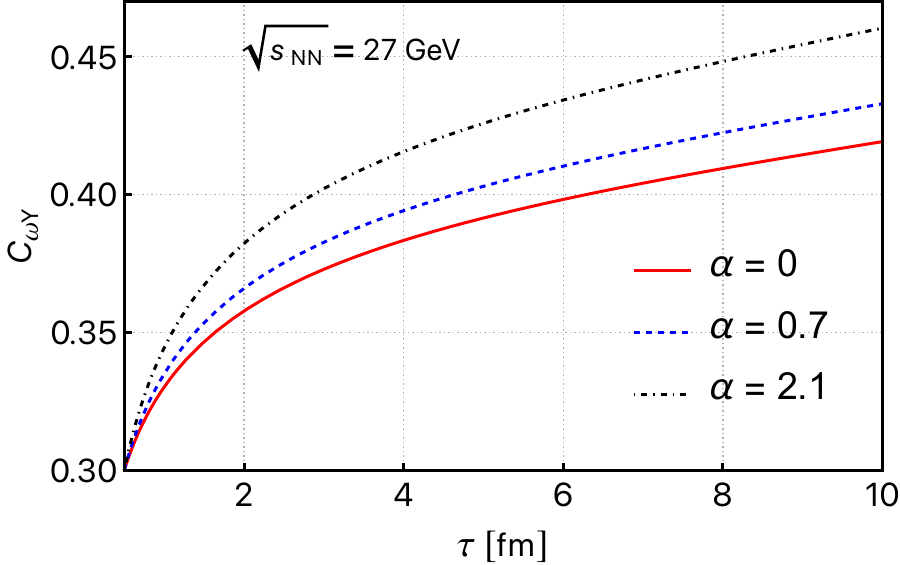}
\caption{Proper-time evolution of spin polarization component $C_{\omega Y}$ for $T_0=0.2$ GeV and $\mu_0 = 0.4$ GeV (upper panel), and $T_0= \mu_0 = 0.3$ GeV (lower panel) corresponding to $\sqrt{s_{\rm NN}} = 7.7$ GeV and $\sqrt{s_{\rm NN}} = 27$ GeV collision energies, respectively. Here, the spin component is initialized as $C_{\omega Y_0} = 0.3$ and $\alpha$ is the parameter to switch on and off the magnetic field where $\alpha = 0$ means system without magnetic field.}
\label{fig:CwY}
\end{figure}
\begin{figure}[!ht]
\centering
\includegraphics[angle=0,width=\columnwidth]{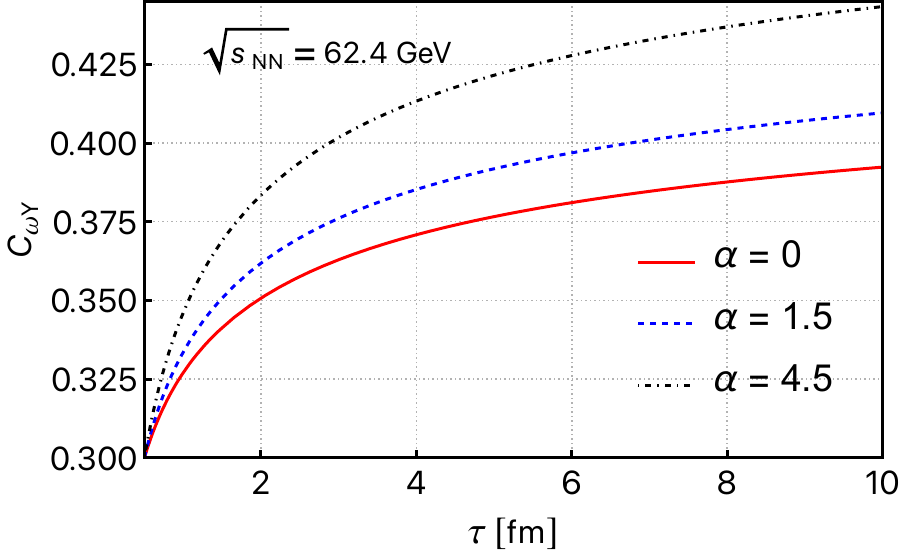}
\includegraphics[angle=0,width=\columnwidth]{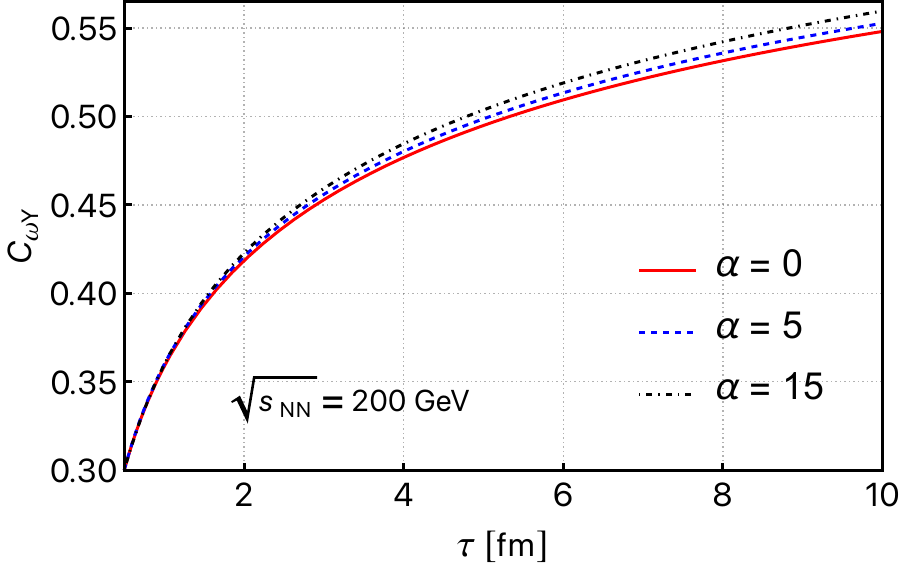}
\caption{Same as Fig.~\ref{fig:CwY} but for $T_0=0.3$ GeV and $\mu_0 = 0.15$ GeV (upper panel), and $T_0 = 0.6$ GeV and $\mu_0 = 0.05$ GeV (lower panel) corresponding to $\sqrt{s_{\rm NN}} = 62.4$ GeV and $\sqrt{s_{\rm NN}} = 200$ GeV collision energies, respectively.}
\label{fig:CwY1}
\end{figure}
We solve the coupled set of equations \rfn{eq:BjorkenCharge}, \rfn{eq:BjorkenEnergy}, and \rfn{eq:SPIN} to obtain the evolution of temperature $T$,
baryon chemical potential $\mu$, and spin polarization components. 
In this setup, each spin polarization component evolve independently of each other~\cite{Florkowski:2019qdp}.
Due to our magnetic field profile~\eqref{eq:magnetic_field}, the evolution of the spin component $C_{\omega Y}$ is modified by the spin-EM coupling, while other components are not affected, thus we will only study the evolution of $C_{\omega Y}$.
As we have assumed small polarization limit, $\omega_{\mu\nu} \ll 1$, the spin component must be initialized accordingly, i.e., it can take any positive value less than 1. Therefore, we choose initial value of the spin component as $C_{\omega Y_0} = 0.3$ at $\tau_0 = 0.5 \, \mathrm{fm}/\mathrm{c}$.
Let us consider four different cases of initial temperature $T_0$ and baryon chemical potential $\mu_0$ corresponding to four different collision energies such as~\cite{Karpenko:2015xea}
\begin{align}
T_0 = 0.2 \, \mathrm{GeV}, \quad \mu_0 = 0.4 \, \mathrm{GeV} \quad {\rm for} \quad \sqrt{s_{\rm NN}} = 7.7 \, {\rm GeV}\,, \nn\\
T_0 = 0.3 \, \mathrm{GeV}, \quad \mu_0 = 0.3 \, \mathrm{GeV} \quad {\rm for} \quad \sqrt{s_{\rm NN}} = 27 \, {\rm GeV}\,, \nn\\
T_0 = 0.3 \, \mathrm{GeV}, \quad \mu_0 = 0.15 \, \mathrm{GeV} \quad {\rm for} \quad \sqrt{s_{\rm NN}} = 62.4 \, {\rm GeV}\,, \nn\\
T_0 = 0.6 \, \mathrm{GeV}, \quad \mu_0 = 0.05 \, \mathrm{GeV} \quad {\rm for} \quad \sqrt{s_{\rm NN}} = 200 \, {\rm GeV}\,,\nn\\
\label{eq:initial_parameters}
\end{align}
to observe the dependence of spin-EM coupling on different initial values of $T$ and $\mu$.

The spin component $C_{\omega Y}$ also has physical relevance with respect to ultra-relativistic heavy-ion collisions, as it is expected that
the total angular momentum is initially purely orbital, negative and directed orthogonal to the reaction plane (i.e. along $-y$ axis)~\cite{STAR:2017ckg,Adam:2018ivw,STAR:2019erd,Acharya:2019ryw,Kornas:2019}.
After the collision, this initial orbital angular momentum can be converted to spin angular momentum in the same direction. Therefore, to address such a physical system it is enough to assume a non-zero and small initial $C_{\omega Y}$ and set all other spin components to zero, see Ref.~\cite{Florkowski:2019qdp} for more details.

Figure~\ref{fig:CwY} shows the evolution of $C_{\omega Y}$ for two different sets of $T_0$ and $\mu_0$ for $\sqrt{s_{\rm NN}} = 7.7$ GeV (upper panel) and $\sqrt{s_{\rm NN}} = 27$ GeV (lower panel), see Eq.~\eqref{eq:initial_parameters}.
Similarly in Fig.~\ref{fig:CwY1} we show the evolution of the spin component $C_{\omega Y}$ for another two sets $T_0$ and $\mu_0$ corresponding to collision energies $\sqrt{s_{\rm NN}} = 62.4$ GeV (upper panel) and $\sqrt{s_{\rm NN}} = 200$ GeV (lower panel).
$\alpha = 0$
corresponds to the system without EM fields~\cite{Florkowski:2019qdp}.
Since $\mu$ is initially positive, the magnetic field intensifies $C_{\omega Y}$ in each case.
This enhancement is transferred to the final polarization of particles. For a fluid with initially negative chemical potential, the magnetic field suppresses the spin polarization. However, such a case is irrelevant to heavy-ion physics. For low collision energies, an increase in the collision energy, gives rise to a decrease in the amplitude of the spin component even in the case of zero magnetic field, see Figs.~\ref{fig:CwY} and \ref{fig:CwY1}. However, when transiting to the so-called high collision energies, for which $\mu_0$ is negligible with respect to $T_0$, the effect of increasing the collision energy is conversed, i.e, it leads to an increase in the spin component's amplitude. For example, see the transition from $\sqrt{s_{\rm NN}} = 62.4$ GeV to $200$ GeV in Fig.~\ref{fig:CwY1}.
Not that we use large values of $\alpha$ for demonstrative purposes. However, qualitative features remain same even for very small values of $\alpha$.
 
The evolution of the spin component $(C_{\omega Y})$ is directly connected to the spin polarization of the $\Lambda(\Bar{\Lambda})$ hyperon emitted at the freeze-out hypersurface, $\Delta \Sigma_{\lambda} = \tau\,U_{\lambda }\,dx\,dy\,d\eta$, through the following expression of average spin polarization per particle ${\langle\pi_{\mu}\rangle}_p$ (suffix $p$ signifies it is a momentum dependent quantity)~\cite{Florkowski:2021wvk,Singh:2022pis}
\begin{equation}
{\langle\pi_{\mu}\rangle}_p = \frac{E_p\frac{d\Pi _{\mu }^{\mbox{*}}(p)}{d^3 p}}{E_p\frac{d{\cal{N}}(p)}{d^3 p}}\,,
\label{meanspin}
\end{equation}
where the numerator is the momentum distribution of the total Pauli-Luba\'nski (PL) four-vector and denominator is the momentum density of particles and antiparticles given, respectively, as~\cite{Florkowski:2021wvk}
 \begin{align}
E_p\frac{d\Pi _{\mu }^{\mbox{*}}(p)}{d^3 p} &= -\f{1}{(2 \pi )^3 m}\Bigg[
\int \cosh(\xi)\,
\Delta \Sigma _{\lambda } p^{\lambda } \,
e^{-\beta \cdot p} \nn\\
&\times \left(\left(\widetilde{\omega }_{\mu \beta }+\widetilde{F}_{\mu \beta }\right)p^{\beta}\right)^{\mbox{*}} \Bigg] ,
\lab{PDLT}
 \end{align}
and
 \begin{equation}
E_p\frac{d{\cal{N}}(p)}{d^3 p}=
\f{2}{(4 \pi )^3}
\int \cosh(\xi)\, \Delta  \Sigma _{\lambda } p^{\lambda } \,e^{-\beta \cdot p}
\,.
\label{MD}
 \end{equation}
Here, $\widetilde{\omega }^{\mu \nu }= (1/2) \epsilon^{\mu\nu\alpha\beta}  {\omega}_{\alpha\beta}$ and $\widetilde{F }^{\mu \nu }= (1/2) \epsilon^{\mu\nu\alpha\beta} {F}_{\alpha\beta}$ are the dual forms of ${\omega }^{\mu \nu }$ and ${F}^{\mu \nu }$, respectively, whereas the asterisk represents that quantities are in the particle rest frame (PRF).
Integrating over momentum coordinates we obtain momentum independent spin polarization using Eq.~\eqref{meanspin} as~\cite{Florkowski:2021wvk}
 \begin{equation}
\langle\pi_{\mu}\rangle ~=~ \frac{\int \mathrm{dP}\, {\langle\pi_{\mu}\rangle}_p \, E_p\frac{d{\cal{N}}(p)}{d^3 p}}{\int \mathrm{dP}\, E_p\frac{d{\cal{N}}(p)}{d^3 p}} ~\equiv~ \frac{\int d^3 p ~\frac{d\Pi_{\mu }^{\mbox{*}}(p)}{d^3 p}}{\int d^3 p~\frac{d{\cal{N}}(p)}{d^3 p}}.
\label{AvgPol}
 \end{equation}
Since Bjorken symmetries force the hydrodynamic variables to be the functions of $\tau$ only, the factor $\cosh(\xi)$, in Eq.~\eqref{meanspin}, comes out of the integration and cancels out~\cite{Florkowski:2019qdp}.
This makes Eq.~\eqref{meanspin} independent of the baryon chemical potential and therefore, within the Bjorken setup, we cannot observe the explicit $\mu$ dependency of $\Lambda-\Bar{\Lambda}$ spin polarization. However, in more realistic setups~\cite{Florkowski:2021wvk,Shi:2022iyb} this might not be the case. Since we assume that magnetic fields do not break Bjorken symmetries, the spin polarization evolution remains independent of the chemical potential with $\alpha\neq 0$.
Thus, we refrain to show them here as they are qualitatively similar to the spin polarization results for Bjorken background without electromagnetic fields~\cite{Florkowski:2019qdp}.
\section{Summary}
\label{sec:summary}
In this work, we introduced a spin-EM coupling term into the classical phase-space distribution function and derived the charge current, energy-momentum tensor, and spin tensor. In contrast to previous studies~\cite{Florkowski:2019qdp,Singh:2020rht,Singh:2021man}, the background and spin equations of motion are highly coupled due to the spin-EM coupling. We also mentioned certain special cases relevant to ultra-relativistic heavy-ion collisions. Lastly, we examined our formalism with Bjorken flow for four different sets of initial temperature and baryon chemical potential and found that with increasing collisional energy, the amplitude of spin component decreases in the low collision energy regime, while it increases otherwise.
With some further considerations, the presented formalism may be relevant in explaining the  observed $\Lambda-\Bar{\Lambda}$ spin polarization splitting. In particular, it is required to consider more realistic setups, which is the focus of our future investigation.

\medskip
{\it Acknowledgements.}
We thank R. Ryblewski for constructive feedback and critical reading of the manuscript and also thank V. E. Ambrus, S. Bhadury, A. Dash, W. Florkowski, A. Jaiswal, D. Rischke, N. Sadooghi, and G. Sophys for fruitful discussions.
M.S. acknowledges support by the Deutsche Forschungsgemeinschaft (DFG, German Research Foundation) through the CRC-TR 211 `Strong-interaction matter under extreme conditions'– project number 315477589 – TRR 211. R.S. acknowledges the support of NAWA PROM Program no: PROM PPI/PRO/2019/1/00016/U/001 and the hospitality of the Institute for Theoretical Physics, Goethe University Germany where part of this work was completed. R.S. also acknowledges the support of Polish NAWA Bekker program no.: BPN/BEK/2021/1/00342 in the completion of this work.
This research was also supported in part by the Polish National Science Centre Grants No. 2016/23/B/ST2/00717 and No. 2018/30/E/ST2/00432.
\appendix
\section{Thermodynamic identities}
\label{app:identities}
In this appendix, we list the thermodynamic identities we used in Sec.~\ref{sec:cons_relations}
 \begin{align}
I_{(0)}^{\mu \nu \lambda \rho} &= I_{40}^{(0)} U^{\mu} U^{\nu} U^{\lambda} U^{\rho} +
I_{41}^{(0)} \big(\Delta^{\mu \nu} U^{\lambda} U^{\rho}
+ \Delta^{\mu \lambda} U^{\nu} U^{\rho}\nn\\
&+ \Delta^{\nu \lambda} U^{\mu} U^{\rho} + \Delta^{\mu \rho} U^{\nu} U^{\lambda} + \Delta^{\nu \rho} U^{\mu} U^{\lambda} + \Delta^{\lambda \rho} U^{\mu} U^{\nu}\big)\nn\\
&+ I_{42}^{(0)} \big(\Delta^{\mu \nu} \Delta^{\lambda \rho} + \Delta^{\mu \lambda} \Delta^{\nu \rho} + \Delta^{\mu \rho} \Delta^{\nu \lambda}\big)\,, \label{eq:I3}
 \end{align}
with
 \begin{align}
I_{40}^{(0)} &= \frac{T^6 z^6}{64  m^2 \pi^2} \left[K_{6}(z) + 2K_{4}(z) - K_{2}(z) - 2K_{0}(z)\right],\\
    I_{41}^{(0)} &= - \frac{T^6 z^6}{192 m^2 \pi^2} \left[K_{6}(z) - 2K_{4}(z) - K_{2}(z) + 2K_{0}(z)\right],\nn\\
    I_{42}^{(0)} &= \frac{T^6 z^6}{960  m^2 \pi^2} \left[K_{6}(z) - 6K_{4}(z) + 15K_{2}(z) - 10K_{0}(z)\right].\nn
 \end{align}
\bibliographystyle{elsarticle-num}
\bibliography{pv_ref}
\end{document}

%% file: commands.tex
\newcommand{\aux}{\color{blue}}

\def\be{\begin{equation}}
	\def\ee{\end{equation}}
	\newcommand{\eel}{\end{eqnarray}}
\def\barr{\begin{array}}
	\def\earr{\end{array}}
\def\beq{\begin{eqnarray}}
	\def\eeq{\end{eqnarray}}
\def\bfig{\begin{figure}}
	\def\efig{\end{figure}}
\newcommand{\bea}{\begin{eqnarray}}
	\newcommand{\eea}{\end{eqnarray}}

\def\LB{\left(}
\def\RB{\right)}
\def\LSB{\left[}
\def\RSB{\right]}
\def\LAB{\langle}
\def\RAB{\rangle}

\newcommand{\VP}{\vphantom{\frac{}{}}\!}
\def\lt{\left}
\def\rt{\right}
\def\CHI{\chi}
\newcommand{\nn}{\nonumber}
\newcommand{\f}[2]{\frac{#1}{#2}}
\newcommand{\onehalf}{{\nicefrac{1}{2}}}
\newcommand{\onethird}{{\nicefrac{1}{3}}}
\newcommand{\fivetwo}{{\nicefrac{5}{2}}}
\newcommand{\p}{\partial}
\newcommand{\del}{\partial}
\newcommand{\rf}[1]{Eq.~(\ref{#1})}
\newcommand{\rfm}[1]{Eqs.~(\ref{#1})}
\newcommand{\rftwo}[2]{Eqs.~(\ref{#1})~and~(\ref{#2})}
\newcommand{\rfmtwo}[2]{Eqs.~(\ref{#1})-(\ref{#2})}
\newcommand{\rfn}[1]{(\ref{#1})}
\newcommand{\rfs}[1]{Sec.~\ref{#1}}
\newcommand{\rff}[1]{Fig.~\ref{#1}}
\newcommand{\rfc}[1]{Ref.~\cite{#1}}
\newcommand{\rfcs}[1]{Refs.~\cite{#1}}
\newcommand{\red}{\color{red}}
\newcommand{\blue}{\color{blue}}
\newcommand{\green}{\color{green}}

\def\a{\alpha}
\def\b{\beta}
\def\g{\gamma}
\def\d{\delta} 
\def\r{\rho}
\def\s{\sigma}
\def\c{\chi} 
 \def\lam{\lambda} 
\def\LR{\left(} 
\def\RR{\right)}
\def\LS{\left[} 
\def\RS{\right]}
\def\LC{\left{} 
\def\RC{\right}}
\def\LA{\left\langle}
\def\RA{\right\rangle}
\def\LD{\left.}
\def\RD{\right.}
\newcommand{\dotb}[1]{\dot{\llbracket} #1 \rrbracket} 
\def\HP{\hphantom{\alpha}} 


\newcommand{\sh}[1]{\sinh#1}
\newcommand{\ch}[1]{\cosh#1}
\newcommand{\shb}[1]{\sinh\LR#1\RR}
\newcommand{\chb}[1]{\cosh\LR#1\RR}
\newcommand{\tU}{\theta_U}
\newcommand{\tX}{\theta_X}
\newcommand{\tY}{\theta_Y}
\newcommand{\tZ}{\theta_Z}
\newcommand{\tI}[1]{\theta_{#1}}
\newcommand{\dU}{d_U}
\newcommand{\dX}{d_X}
\newcommand{\dY}{d_Y}
\newcommand{\dZ}{d_Z}
\newcommand{\dI}[1]{d_{#1}}

\def\half{\frac{1}{2}}

\def\GLW{{\rm GLW}}
\def\LRFF{{\rm LRFF}}


\def\nU{n_{(0)}}
\def\nUi{n_{(0),i}}
\def\eU{\varepsilon_{(0)}}
\def\eUi{\varepsilon_{(0),i}}
\def\PU{P_{(0)}}
\def\PUi{P_{(0),i}}
\def\sU{s_{(0)}}
\def\sU{s_{(0),i}}

\def\nP{n_{}}
\def\eP{\varepsilon_{}}
\def\PP{P_{}}
\def\sP{s_{}}
\def\wP{w_{}}

\newcommand{\lab}[1]{\label{#1}}
\def\nn{\nonumber}

\newcommand{\refb}[1]{(\ref{#1})}
\newcommand{\refeq}[1]{Eq.~(\ref{#1})}
\newcommand{\refeqs}[1]{Eqs.~(\ref{#1})}


\def\cA{{\cal A}}
\def\cB{{\cal B}}
\def\cC{{\cal C}}
\def\cD{{\cal D}}
\def\cN{{\cal N}}
\def\cE{{\cal E}}
\def\cP{{\cal P}}
\def\cS{{\cal S}}
\def\cT{{\cal T}}
\def\cQ{{\cal Q}}
\def\cNN{{\cal N}_{(0)}}
\def\cEN{{\cal E}_{(0)}}
\def\cPN{{\cal P}_{(0)}}
\def\cSN{{\cal S}_{(0)}}


\def\pmu{p^\mu}
\def\pnu{p^\nu}

\def\vv{{\boldsymbol v}}
\def\pv{{\boldsymbol p}}
\def\av{{\boldsymbol a}}
\def\bv{{\boldsymbol b}}
\def\kv{{\boldsymbol k}}
\def\omnL{\omega_{\mu\nu}}
\def\omnU{\omega^{\mu\nu}}
\def\omnLbar{{\bar \omega}_{\mu\nu}}
\def\omnUbar{{\bar \omega}^{\mu\nu}}
\def\omnLbardot{{\dot {\bar \omega}}_{\mu\nu}}
\def\omnUbardot{{\dot {\bar \omega}}^{\mu\nu}}

\def\oabL{\omega_{\alpha\beta}}
\def\oabU{\omega^{\alpha\beta}}
\def\omnLD{{\tilde \omega}_{\mu\nu}}
\def\omnUD{\tilde {\omega}^{\mu\nu}}
\def\omnLDbar{{\bar {\tilde \omega}}_{\mu\nu}}
\def\omnUDbar{{\bar {\tilde {\omega}}}^{\mu\nu}}
\def\CHI{\chi}
\def\bchem{\mu_{\rm B}}
\def\bfug{\xi_{\rm B}}
\def\tfug{\xi}

\def\be{\begin{equation}}
\def\ee{\end{equation}}
\def\ba{\begin{eqnarray}}
\def\ea{\end{eqnarray}}   

\def\a{\alpha}
\def\b{\beta}
\def\g{\gamma}
\def\d{\delta} 
\def\r{\rho}
\def\s{\sigma}
\def\c{\chi}
 
\def\LR{\left(} 
\def\RR{\right)}
\def\LS{\left[} 
\def\RS{\right]}
\def\LC{\left{} 
\def\RC{\right}}
\def\LA{\left\langle}
\def\RA{\right\rangle}
\def\LD{\left.}
\def\RD{\right.}
\def\half{\frac{1}{2}}

\def\GLW{{\rm GLW}}
\def\LRF{{\rm LRF}}


\def\nU{n_{(0)}}
\def\eU{\varepsilon_{(0)}}
\def\PU{P_{(0)}}
\def\sU{s_{(0)}}

\def\nP{n_{}}
\def\eP{\varepsilon_{}}
\def\PP{P_{}}
\def\sP{s_{}}
\def\wP{w_{}}


\def\pmu{p^\mu}
\def\pnu{p^\nu}

\def\vv{{\boldsymbol v}}
\def\pv{{\boldsymbol p}}
\def\av{{\boldsymbol a}}
\def\bv{{\boldsymbol b}}
\def\cv{{\boldsymbol c}}
\def\Cv{{\boldsymbol C}}
\def\kv{{\boldsymbol k}}
\def\piv{{\boldsymbol \pi}}

\def\thetap{\theta_\perp}
\def\omnL{\omega_{\mu\nu}}
\def\omnU{\omega^{\mu\nu}}
\def\omnLbar{{\bar \omega}_{\mu\nu}}
\def\omnUbar{{\bar \omega}^{\mu\nu}}
\def\omnLbardot{{\dot {\bar \omega}}_{\mu\nu}}
\def\omnUbardot{{\dot {\bar \omega}}^{\mu\nu}}

\def\oabL{\omega_{\alpha\beta}}
\def\oabU{\omega^{\alpha\beta}}
\def\omnLD{{\tilde \omega}_{\mu\nu}}
\def\omnUD{\tilde {\omega}^{\mu\nu}}
\def\omnLDbar{{\bar {\tilde \omega}}_{\mu\nu}}
\def\omnUDbar{{\bar {\tilde {\omega}}}^{\mu\nu}}

\def\epsLmnbg{\epsilon_{\mu\nu\beta\gamma}}
\def\epsUmnbg{\epsilon^{\mu\nu\beta\gamma}}
\def\epsLmnab{\epsilon_{\mu\nu\alpha\beta}}
\def\epsUmnab{\epsilon^{\mu\nu\alpha\beta}}

\def\epsUmnrs{\epsilon^{\mu\nu\rho \sigma}}
\def\epsUlnrs{\epsilon^{\lambda \nu\rho \sigma}}
\def\epsUlmrs{\epsilon^{\lambda \mu\rho \sigma}}

\def\epsLmnbg{\epsilon_{\mu\nu\beta\gamma}}
\def\epsUmnbg{\epsilon^{\mu\nu\beta\gamma}}
\def\epsLmnab{\epsilon_{\mu\nu\alpha\beta}}
\def\epsUmnab{\epsilon^{\mu\nu\alpha\beta}}

\def\epsLabgd{\epsilon_{\alpha\beta\gamma\delta}}
\def\epsUabgd{\epsilon^{\alpha\beta\gamma\delta}}

\def\epsUmnrs{\epsilon^{\mu\nu\rho \sigma}}
\def\epsUlnrs{\epsilon^{\lambda \nu\rho \sigma}}
\def\epsUlmrs{\epsilon^{\lambda \mu\rho \sigma}}

\def\epsLijk{\epsilon_{ijk}}


\def\epsLmnbg{\epsilon_{\mu\nu\beta\gamma}}
\def\epsUmnbg{\epsilon^{\mu\nu\beta\gamma}}
\def\epsLmnab{\epsilon_{\mu\nu\alpha\beta}}
\def\epsUmnab{\epsilon^{\mu\nu\alpha\beta}}

\def\epsUmnrs{\epsilon^{\mu\nu\rho \sigma}}
\def\epsUlnrs{\epsilon^{\lambda \nu\rho \sigma}}
\def\epsUlmrs{\epsilon^{\lambda \mu\rho \sigma}}

\def\epsLmnbg{\epsilon_{\mu\nu\beta\gamma}}
\def\epsUmnbg{\epsilon^{\mu\nu\beta\gamma}}
\def\epsLmnab{\epsilon_{\mu\nu\alpha\beta}}
\def\epsUmnab{\epsilon^{\mu\nu\alpha\beta}}

\def\epsLabgd{\epsilon_{\alpha\beta\gamma\delta}}
\def\epsUabgd{\epsilon^{\alpha\beta\gamma\delta}}

\def\epsUmnrs{\epsilon^{\mu\nu\rho \sigma}}
\def\epsUlnrs{\epsilon^{\lambda \nu\rho \sigma}}
\def\epsUlmrs{\epsilon^{\lambda \mu\rho \sigma}}

\def\epsLijk{\epsilon_{ijk}}
\def\half{\frac{1}{2}}
\def\GLW{{\rm GLW}}

\def\n0{n_{(0)}}
\def\e0{\varepsilon_{(0)}}
\def\P0{P_{(0)}}
\newcommand{\redflag}[1]{{\color{red} #1}}
\newcommand{\blueflag}[1]{{\color{blue} #1}}
\newcommand{\checked}[1]{{\color{darkblue} \bf [#1]}}
\newcommand{\Psis}{{\sf \Psi}}
\newcommand{\psis}{{\sf \psi}}
\newcommand{\Psibar}{{\overline \Psi}}
\def\eMf{electromagnetic (EM) }
\def\EMf{Electromagnetic (EM) }
\def\EM{EM }
\def\lRFf{local rest frame (LRF)}
\def\LRFf{Local rest frame (LRF) }
\def\LRF{LRF }
\def\QGPf{Quark gluon plasma (QGP) }
\def\qGPf{Quark gluon plasma (QGP) }
\def\QGP{QGP }
\def\mHDf{magnetohydrodynamic (MHD) }
\def\MHDf{Magnetohydrodynamic (MHD) }
\def\MHD{MHD }
\def\iMHD{iMHD }
\def\HD{Hydrodynamics }
\def\hD{hydrodynamics }
\def\RHD{Relativistic hydrodynamics }
\def\rHD{relativistic hydrodynamics }
\def\rMHDf{relativistic magnetohydrodynamic (RMHD) }
\def\RMHDf{Relativistic magnetohydrodynamic (RMHD) }
\def\RMHD{RMHD }
\def\eOMf{equations of motion (EOM)~}
\def\EOMf{Equations of motion (EOM)~}
\def\EOM{EOM}
\def\fl{\ensuremath{\text{Fluid}}}
\def\lrf{\ensuremath{\text{LRF}}}
\def\BVf{Boltzmann-Vlasov (BV) }
\def\BV{BV\,}
		
\def\rhoLEQ{{\widehat{\rho}}_{\rm \small LEQ}}
\def\rhoGEQ{{\widehat{\rho}}_{\rm \small GEQ}}
		
\def\fplushat{{\hat f}^+}
\def\fminushat{{\hat f}^-}
		
\def\fplusrs{f^+_{rs}}
\def\fplussr{f^+_{sr}}
\def\fplusrsxp{f^+_{rs}(x,p)}
\def\fplussrxp{f^+_{sr}(x,p)}
		
\def\fminusrs{f^-_{rs}}
\def\fminussr{f^-_{sr}}
\def\fminusrsxp{f^-_{rs}(x,p)}
\def\fminussrxp{f^-_{sr}(x,p)}
		
\def\fpmrs{f^\pm_{rs}}
\def\fpmrsxp{f^\pm_{rs}(x,p)}

\def\feqplus{f^+_{eq}}
\def\feqplus{f^+_{eq}}
\def\feqplusxp{f^+_{eq}(x,p)}
\def\feqplusxp{f^+_{eq}(x,p)}
		
\def\feqminus{f^-_{eq}}
\def\feqminus{f^-_{eq}}
\def\feqminusxp{f^-_{eq}(x,p)}
\def\feqminusxp{f^-_{eq}(x,p)}
	
\def\feqpm{f^\pm_{{\rm eq}}}
\def\feqpmxp{f^\pm_{{\rm eq}}(x,p)}
\def\feqpmi{f^\pm_{{\rm eq},i}}
\def\feqpmxpi{f^\pm_{{\rm eq},i}(x,p)}
\def\fpm{f^\pm}
\def\fpmxp{f^\pm(x,p)}
\def\fpmi{f^\pm_i}
\def\fpmxpi{f^\pm_i(x,p)}
\newcommand{\rs}[1]{\textcolor{red}{#1}}
\newcommand{\rrin}[1]{\textcolor{blue}{#1}}
\newcommand{\rrout}[1]{\textcolor{blue}{\sout{#1}}}
\newcommand{\lie}[2]{\pounds_{#1}\,#2}
\newcommand{\rd}{\mathrm{d}}
\def\re{\mathrm{e}}
\def\echarge{\ensuremath{\rho_e}}
\def\cond{\ensuremath{{\sigma_e}}}
\newcommand{\msnote}[1]{\todo[author=Masoud]{#1}}
\newcommand{\msnotei}[1]{\todo[author=Masoud,inline]{#1}}
\newcommand{\explainindetail}[1]{\todo[color=red!40]{#1}}
\newcommand{\insertref}[1]{\todo[color=green!40]{#1}}
\newcommand{\fm}{\rm{\,fm}}
\newcommand{\fmc}{\rm{\,fm/c}}


\def\uv{{\boldsymbol U}}


\def\kbarzero{ {\bar k}^0}
\def\kv{{\boldsymbol k}}
\def\kbarv{{\bar {\boldsymbol k}}}

\def\obarzero{ {\bar \omega}^0}
\def\ov{{\boldsymbol \omega}}
\def\obar{{\bar \omega}}
\def\obarv{{\bar {\boldsymbol \omega}}}

\def\ev{{\boldsymbol e}}
\def\bv{{\boldsymbol b}}
\newcommand{\tT}{\theta_T}
\newcommand{\UD}[1]{\oU{#1}}
\newcommand{\XD}[1]{\oX{#1}}
\newcommand{\YD}[1]{\oY{#1}}
\newcommand{\ZD}[1]{\oZ{#1}}
\def\Aone{{ \cal A}_1 }
\def\Atwo{{ \cal A}_2 }
\def\Athree{{ \cal A}_3 }
\def\Afour{{ \cal A}_4 }
\def\vv{{\boldsymbol v}}
\def\pv{{\boldsymbol p}}

\newcommand{\inv}[1]{\frac{1}{#1}}
\newcommand{\iinv}[1]{1/#1}